\documentclass[12pt]{iopart}
\usepackage{cite}
\begin{document}
\title[Strong subadditivity condition for qudit state]
{Quantum strong subadditivity condition for systems without subsystems}
\author{Margarita A Man'ko and Vladimir I Man'ko}
\address{P N Lebedev Physical Institute, Leninskii Prospect 53,
Moscow 119991, Russia} \ead{mmanko@sci.lebedev.ru} \ead{manko@sci.lebedev.ru}
\begin{abstract}
The strong subadditivity condition for the density matrix of a quantum system,
which does not contain subsystems, is derived using the qudit-portrait method.
An example of the qudit state in the seven-dimensional Hilbert space
corresponding to spin $j=3$ is presented in detail. New entropic inequalities
in the form of subadditivity condition and strong subadditivity condition for
spin tomograms determining the qudit states are obtained and given on examples
of $j=2$ and 3.
\end{abstract}

\pacs{03.65.-w, 03.65.Ta, 02.50.Cw, 03.67.-a}








\section{Introduction}
The quantum correlations between the subsystems of composite systems provide
specific entropic inequalities relating von Neumann entropies of the system
and its subsystems. For example, the quantum correlations are responsible for
violation of classical entropic inequality for bipartite classical systems
$H(1)\leq H(1,2)$, where $H(1,2)$ is the Shannon entropy of bipartite
classical system and $H(1)$ is the Shannon entropy\cite{Shannon} of its
subsystem. This inequality having intuitively clear interpretation that the
disorder in total system is either the same or larger than the disorder of its
subsystems is not true for quantum bipartite system.

It is known that, for two-qubit pure maximum entangled state with density
matrix $\rho(1,2)$, the von Neumann entropy
$S(1,2)=-\mbox{Tr}\,\rho(1,2)\ln\rho(1,2)=0$, but von Neumann entropy for
one-qubit state $S(1)=-\mbox{Tr}\,\rho(1)\ln\rho(1)$ with
$\rho(1)=-\mbox{Tr}_2\,\rho(1,2)$ has the maximum possible for qubit value,
i.e., $S(1)= \ln 2$. Thus, in this state $S(1)>S(1,2)$, i.e., quantum
correlations between two qubits in the composite system (consisting of two
qubits) provide not only the violation of the Bell
inequalities~\cite{Bell,HornClauser} but also yield the violation of the
classical entropic inequality.

For bipartite systems, both classical and quantum, there exist entropic
inequalities, called the subadditivity conditions, which are the same for
Shannon entropies and von Neumann entropies. The quantum subadditivity
condition can be proved, e.g., by using the tomographic-probability
description of spin states~\cite{Mendes}. Recent review of the tomographic
representation of classical and quantum mechanics can be found in
\cite{NuovoCim,VovaJETP}. For tripartite systems, both classical and quantum,
there also exist entropic inequalities, called the strong subadditivity
conditions, which have the same form for classical Shannon entropies in the
classical case and for von Neumann entropies in the quantum case. Lieb and
Ruskai were the first who proved the quantum strong subadditivity
condition~\cite{LiebRuskai}. The tomographic-probability approach to the
strong subadditivity condition was discussed in \cite{Mendes}. Various aspects
of entropic inequalities and the quantum strong subadditivity condition for
three-partite systems can be found in \cite{Ruskai,Lieb,6,4,5,8,RitaFP}.

Recently, it was shown that the subadditivity condition exists not only in
bipartite quantum systems but also in the systems which do not contain
subsystems, e.g., for one qutrit~\cite{VovJRLR2013}. An approach to derive the
subadditivity condition for the qutrit state is based on the method called the
qubit portrait of qudit states~\cite{portJRLR}, later on used in
\cite{LupoJPA} to study the entanglement in two qudit systems.

The aim of this paper is to show that the strong subadditivity condition can
be obtained for the quantum systems which do not have subsystems. For this, we
apply the qudit-portrait method (which is a generalization of the
qubit-portrait method) that, in fact, is acting by a specific positive map on
the density matrix. The map is described by the action on a vector by the
matrix with matrix elements equal either to zero or unity. Such matrices are
used to get density matrices of subsystems by partial tracing of the density
matrices of the composite-system states. In this case, the system density
matrix first is mapped onto the vector, and then the map matrix acts onto this
vector.  The obtained vector is mapped again onto the new density matrix. For
composite systems, the portrait method is identical to taking the partial
trace of the system density matrix. Since the $N$-dimensional density matrix
of the composite system state and the state of a qudit in the $N$-dimensional
Hilbert space have the identical properties, there exists a possibility to
obtain and apply the strong subadditivity condition available for the
composite system to the system without subsystems.

Our aim is to present the strong subadditivity condition for an arbitrary
probability $N$-vector describing the classical system without subsystems. In
the quantum case, we present the strong subadditivity condition for an
arbitrary density $N$$\times$$N$-matrix describing the state of a system
without subsystems.

This paper is organized as follows.

In section~2, we consider the classical system described by the probability
$N$-vector and show the example of $N=7$ in detail. In section~3, we consider
the quantum system state associated with the density $N$$\times$$N$-matrix and
present the example of $N=7$. We give our conclusions and prospectives in
section~4, where we also discuss the possible consequences for the systems of
qudits and quantum correlations in these systems in the context of strong
subadditivity conditions obtained. In Appendix, the entropic inequalities for
tomograms of some qudit states are presented.

\section{Classical strong subadditivity condition}
We consider a classical system for which one has a random variable. The
probabilities to get the values of this random variable are described by a
probability vector $\vec p=(p_1,p_2,\ldots,p_N)$, where $p_k\geq 0$ and
$\sum_{k=1}^Np_k=1$. The system has no subsystems, and the order in this
system is described by the Shannon entropy
\begin{equation}\label{1}
H=-\sum_{k=1}^Np_k\ln p_k,
\end{equation}
which satisfies the inequality $H\geq 0$, takes the maximum value for $\vec p$
with components $p_k=N^{-1}$, and equals $H_{\rm max}=\ln N$.

If the classical system has two subsystems 1 and 2 and two random variables,
the probability to get the two values of these random variables is described
by the nonnegative numbers ${\cal P}_{kj}$, $k=1,2,\ldots, N_1$, and
$j=1,2,\ldots, N_2$. The probabilities satisfy the normalization condition
$\sum_{k=1}^{N_1}\sum_{j=1}^{N_2}{\cal P}_{kj}=1$, and the Shannon entropy of
the system state reads
\begin{equation}\label{2}
H(1,2)=-\sum_{k=1}^{N_1}\sum_{j=1}^{N_2}{\cal P}_{kj}\ln{\cal P}_{kj}.
\end{equation}
The joint probability distribution ${\cal P}_{kj}$ provides the marginal
distributions for systems 1 and 2 as follows:
\begin{equation}\label{3}
P_{1k}=\sum_{j=1}^{N_2}{\cal P}_{kj},\quad P_{2j}=\sum_{k=1}^{N_1}{\cal
P}_{kj}.
\end{equation}
Thus, we have two Shannon entropies associated with marginal
distributions~(\ref{3}), and they read
\begin{equation}\label{4}
H(1)=-\sum_{k=1}^{N_1}P_{1k}\ln P_{1k},\quad H(2)=-\sum_{j=1}^{N_2} P_{2j}\ln
P_{2j}.
\end{equation}
It is known that these entropies satisfy the subadditivity condition written
in the form of inequality
\begin{equation}\label{5}
H(1)+H(2)\geq H(1,2),
\end{equation}
and the Shannon information is defined as the difference
\begin{equation}\label{6}
I=H(1)+H(2)-H(1,2).
\end{equation}

If the classical system has three subsystems (1, 2, and 3) with three random
variables, the joint probability distribution describing the results of
measurement of the random variables is related to the nonnegative numbers
$\Pi_{kjl}$ $(k=1,2,\ldots N_1$, $j=1,2,\ldots N_2$, and $l=1,2,\ldots N_3)$.
The nonnegative numbers determine the marginal probability distributions
\begin{equation}\label{7}
 {\cal P}_{kj}^{(12)}=\sum_{l=1}^{N_3}\Pi_{kjl},\quad{\cal
P}_{jl}^{(23)}=\sum_{k=1}^{N_1}\Pi_{kjl},\quad 
P_{j}^{(2)}=\sum_{k=1}^{N_1}\sum_{l=1}^{N_3}\Pi_{kjl}.
\end{equation}
The Shannon entropies associated with these probability distributions satisfy
the strong subadditivity condition
\begin{equation}\label{9}
H(1,2)+H(2,3)\geq H(1,2,3)+H(2),
\end{equation}
where
\begin{equation}\label{10}
H(1,2,3)=-\sum_{k=1}^{N_1}\sum_{j=1}^{N_2}\sum_{l=1}^{N_3}\Pi_{kjl}\ln\Pi_{kjl},
\end{equation}
and entropies $H(1,2)$, $H(2,3)$, and $H(2)$ associated with distributions
${\cal P}_{kj}^{(12)}$, ${\cal P}_{jl}^{(23)}$, and $P_{j}^{(2)}$ are given by
(\ref{2}) and (\ref{4}) with obvious substitutions.

In \cite{VovJRLR2013}, it was suggested to obtain an analog of the
subadditivity condition~(\ref{5}) for the system without subsystems. The
general scheme to get such inequality is to write the probability vector $\vec
p$ with components $p_k$ $(k=1,2,\ldots,N)$ in a matrix form with matrix
elements ${\cal P}_{kj}$. Then inequality~(\ref{5}) can be obtained in view of
the above procedure. Here, we apply this method to map the probability vector
$\vec p$ onto the table of numbers with three indices $\Pi_{kjl}$. As a
result, we can obtain the strong subadditivity condition for the system
without subsystems. We demonstrate this procedure on the example of $\vec p$
with 8 components.

Let us define a map given by the equalities
\begin{eqnarray}
&&p_1=\Pi_{111},\quad p_2=\Pi_{112},\quad p_3=\Pi_{121},\quad
p_4=\Pi_{122},\nonumber\\[-2mm]
&&\label{11}\\[-2mm]
&&p_5=\Pi_{211},\quad p_6=\Pi_{212},\quad p_7=\Pi_{221},\quad
p_8=\Pi_{222}.\nonumber
\end{eqnarray}
The map introduced provides an inequality, which is the strong subadditivity
condition associated with the table $\Pi_{kjl}$. To point out a peculiarity of
the strong subadditivity condition, we consider the case of $N=7$. It is the
prime number, and the system with the probability vector has no subsystems.
Thus, we have 7 nonnegative numbers $p_1,p_2,\ldots,p_7$ and the normalization
condition $p_1+p_2+\cdots+p_7=1$. Also we add an extra component $p_8=0$ to
the probability vector. We added zero components to the probability vector
since there is a mismatch of numbers $2^n$ and $2k+1$ (in the case under
consideration, numbers 8 and 7). This means that, in the previous picture of
the 8-dimensional probability vector, we consider the probability distribution
with the constraint $p_8=0$ that provides the constraint $\Pi_{222}=0$ in
map~(\ref{11}).

Applying inequality~(\ref{9}) and formula~(\ref{10}), we obtain the strong
subadditivity condition in the case of the probability 7-vector, which we
express in an explicit form in terms of the vector components
\begin{eqnarray}
\fl\left(-\sum_{k=1}^{7}p_{k}\ln
p_{k}\right)-(p_1+p_2+p_5+p_6)\ln(p_1+p_2+p_5+p_6)\nonumber\\
\fl-(p_3+p_4+p_7)\ln(p_3+p_4+p_7)
\leq -(p_1+p_2)\ln(p_1+p_2)-(p_3+p_4)\ln(p_3+p_4)\nonumber\\
\fl-(p_5+p_6)\ln(p_5+p_6)-p_7\ln
p_7-(p_1+p_5)\ln(p_1+p_5)-(p_2+p_6)\ln(p_2+p_6)\nonumber\\
\fl -(p_3+p_7)\ln(p_3+p_7)-p_4\ln p_4.\label{12}
\end{eqnarray}
This inequality is valid if one makes an arbitrary permutation of $7!$
permutations of the vector components of the probability vector $\vec p$.
Inequality~(\ref{12}) can be presented in the form, where the terms $-p_4\ln
p_4$ and $-p_7\ln p_7$ are removed from the both sides of the inequality.

Analogously, we can write the subadditivity condition following the approach
of \cite{VovJRLR2013}. For example, we have
\begin{eqnarray}
&&-\sum_{k=1}^{7}p_{k}\ln p_{k}\leq
-(p_1+p_2+p_5+p_6)\ln(p_1+p_2+p_5+p_6)\nonumber\\
&&-(p_3+p_4+p_7)\ln(p_3+p_4+p_7)-(p_1+p_3)\ln(p_1+p_3)\nonumber\\
&&-(p_2+p_4)\ln(p_2+p_4)-(p_5+p_7)\ln(p_5+p_7)-p_4\ln p_4.\label{13}
\end{eqnarray}
Also we can rewrite this inequality removing the term $-p_4\ln p_4$ from the
both sides of the inequality. We see that this inequality is valid for a
system without subsystems. For example, in the case of quantum particle with
spin $j=3$, the state of this particle is determined by the spin tomogram
$w(m,\vec n)$~\cite{DodPLA,OlgaJETP}, where the spin projection
$m=-3,-2,-1,0,1,2,3$, and the unit vector $\vec n$ determines the quantization
axes. The tomographic-probability distribution (spin tomogram) of any qudit
state with the density matrix $\rho$ is determined by diagonal matrix elements
of the rotated density matrix as $w(m,\vec n)=\langle m\mid u\rho
u^\dagger\mid m\rangle$, where the unitary matrix $u$ is the matrix of
irreducible representation of the rotation group, and it depends on the Euler
angles determining the unit vector $\vec n$. Thus, the tomogram is the
probability distribution of the spin projection $m$ on the direction $\vec n$.
We can identify the components of the probability vector $\vec p$ with the
tomographic probabilities. Then we have the inequality -- the subadditivity
condition for the spin tomographic probabilities:
\begin{eqnarray}\label{sub1}\fl
-\sum_{ m=-3}^3w(m,\vec n)\ln w(m,\vec n)\leq -\big[w(-3,\vec n)+w(-2,\vec
n)+w(1,\vec n)+w(2,\vec n)\big]\nonumber\\
\hspace{30mm}\times\ln\big[w(-3,\vec n)+w(-2,\vec
n)+w(1,\vec n)+w(2,\vec n)\big]\nonumber\\
-\big[w(-1,\vec n)+w(0,\vec n)+w(3,\vec n)\big]\ln\big[w(-1,\vec n)+w(0,\vec
n)+w(3,\vec
n)\big]\nonumber\\
-\big[w(-3,\vec n)+w(1,\vec n)\big]\ln\big(w(-3,\vec n)+w(1,\vec
n)\big]\nonumber\\ -\big[w(-2,\vec n)+w(2,\vec n)\big)\ln\big[w(-2,\vec
n)+w(2,\vec
n)\big]\nonumber\\
-\big[w(-1,\vec n)+w(3,\vec n)\big]\ln\big[w(-1,\vec n)+w(3,\vec
n)\big]-w(0,\vec n)\ln w(0,\vec n).
\end{eqnarray}
This inequality describes some properties of quantum correlations in the spin
system with $j=3$. In spite of the fact that this system does not have
subsystems, inequality~(\ref{sub1}), being corresponded to the subadditivity
condition, is valid for any direction of the vector $\vec n$. Other examples
of tomographic inequalities are given in Appendix.

\section{Strong subadditivity condition for one qudit state}
In this section, we obtain the strong subadditivity condition for a system
without subsystems written in the form of an inequality for von Neumann
entropies associated with the initial density matrix of the spin-$j$ state and
its qubit (or qudit) portraits. The qubit (or qudit)
portrait~\cite{portJRLR,MVJRLR} of the initial density matrix is a specific
positive map of this matrix obtained following the procedure: Any
$N$$\times$$N$-matrix $\rho_{jk}$ is considered as the column vector
$\vec\rho$ with components $(\rho_{11},\rho_{12},\ldots,\rho_{1N},
\rho_{21},\rho_{22},\ldots,\rho_{2N},\ldots,\rho_{N1},\rho_{N2},\ldots,
\rho_{NN})$. We multiply this vector by the matrix $M$ which contains only
units and zeros, and the units and zeros are matrix elements to provide that
the new vector $\vec\rho_M$ obtained is considered as a new matrix
$(\rho_M)_{jk}$ being the density matrix. It is easy to prove that, for any
density matrix of a multi-qudit system $\rho(1,2,\ldots, M)$, one can
calculate the density matrix of an arbitrary subsystem of qudits
$\rho(1,2,\ldots, M')$ by means of a portrait of the initial density matrix.
In view of this observation, we extend the entropic inequalities available for
composite systems of qudits to arbitrary density matrices including the
density matrices of a single qudit. We show the result of such an approach on
an example of the strong subadditivity condition known for three-partite
quantum systems~\cite{LiebRuskai}.

For the qudit state with $j=3$ and the density matrix $\rho$ with matrix
elements $\rho_{kj}$, $k,j=1,2,\ldots,7$, the strong subadditivity condition
found turns out to be
\begin{equation}\label{SSC1}
-\mbox{Tr}\left(\rho\ln\rho\right)-\mbox{Tr}\left(R_2\ln R_2\right)\leq
-\mbox{Tr}\left(R_{12}\ln R_{12}\right)-\mbox{Tr}\left(R_{23}\ln
R_{23}\right),
\end{equation}
where the density matrix $R_{12}$ has matrix elements expressed in terms of
the density matrix $\rho_{jk}$ as follows:
\begin{equation}\label{SSC12} R_{12}=
\pmatrix{\rho_{11}+\rho_{22}&\rho_{13}+\rho_{24}&\rho_{15}+\rho_{26}&\rho_{17}\cr
\rho_{31}+\rho_{42}&\rho_{33}+\rho_{44}&\rho_{35}+\rho_{46}&\rho_{37}\cr
\rho_{51}+\rho_{62}&\rho_{53}+\rho_{64}&\rho_{55}+\rho_{66}&\rho_{57}\cr
\rho_{71}&\rho_{73}&\rho_{75}&\rho_{77}\cr}.\end{equation} The density matrix
$R_{23}$ reads
\begin{equation}\label{SSC23} R_{23}=
\pmatrix{\rho_{11}+\rho_{55}&\rho_{12}+\rho_{56}&\rho_{13}+\rho_{57}&\rho_{14}\cr
\rho_{21}+\rho_{65}&\rho_{22}+\rho_{66}&\rho_{23}+\rho_{67}&\rho_{24}\cr
\rho_{31}+\rho_{75}&\rho_{32}+\rho_{76}&\rho_{33}+\rho_{77}&\rho_{34}\cr
\rho_{41}&\rho_{42}&\rho_{43}&\rho_{44}\cr},\end{equation} while the matrix
$R_{2}$ is
\begin{equation}\label{SSC2} R_2=
\pmatrix{\rho_{11}+\rho_{22}+\rho_{55}+\rho_{66}&\rho_{13}+\rho_{24}+\rho_{57}\cr
\rho_{31}+\rho_{42}+\rho_{75}&\rho_{33}+\rho_{44}+\rho_{77}\cr}.\end{equation}

The inequality for von Neumann entropies associated with the matrices $\rho$,
$R_{12}$,  $R_{23}$, and  $R_{2}$ has a form of the strong subadditivity
condition for a three-partite system with the density matrix $\rho(1,2,3)$
obtained in \cite{LiebRuskai}.

The other entropic inequality for the spin-2 state with the density matrix
$\rho_{jk}$, $j,k=1,2,3,4,5$ has the form~(\ref{SSC1}) with the matrices
$R_{12}$,  $R_{23}$, and  $R_{2}$ as follows:
\begin{eqnarray}\fl R_{12}=
\pmatrix{\rho_{11}+\rho_{22}&\rho_{13}+\rho_{24}&\rho_{15}\cr
\rho_{31}+\rho_{42}&\rho_{33}+\rho_{44}&\rho_{35}\cr
\rho_{51}&\rho_{53}&\rho_{55}\cr},\qquad R_{23}=
\pmatrix{\rho_{11}+\rho_{55}&\rho_{12}&\rho_{13}&\rho_{14}\cr
\rho_{21}&\rho_{22}&\rho_{23}&\rho_{24}\cr
\rho_{31}&\rho_{32}&\rho_{33}&\rho_{34}\cr
\rho_{41}&\rho_{42}&\rho_{43}&\rho_{44}\cr},\label{SSC12-2}\\  R_2=
\pmatrix{\rho_{11}+\rho_{22}+\rho_{55}&\rho_{13}+\rho_{24}\cr
\rho_{31}+\rho_{42}&\rho_{33}+\rho_{44}\cr}.\label{SSC2-2}
\end{eqnarray}

\section{Conclusions}
To conclude, we list our main results.

We proved matrix inequalities for arbitrary nonnegative Hermitian $N$$\times
$$N$-matrices with trace equal to unity. If the matrix is identified with
the density matrix of qudit state, the matrix inequalities obtained are
entropic inequalities characterizing quantum correlations in the system.

Employing the positive map of an arbitrary density matrix corresponding to the
qubit (or qudit) portrait of the density matrix of a multiqudit state
identified with the calculation of the subsystem-state density matrices, we
obtained an analog of the strong subadditivity condition for the state of the
system, which does not contain any subsystems. This result is an extension of
the approach~\cite{VovJRLR2013}, where the subadditivity condition was
obtained for quantum systems without subsystems. We derived the entropic
inequalities for the qudit-state tomograms and showed examples of the
subadditivity condition and the strong subadditivity condition for the spin
states with $j=2$ and $j=3$, respectively. We presented the entropic
inequalities for density matrices --- analogs of the strong subadditivity
condition for $j=3$
--- in the form of an explicit matrix inequality. We formulated the approach
to find new entropic inequalities for both cases: (i)~the probability
distributions and related Shannon entropies and (ii)~the density matrices and
related von Neumann entropies.

For given arbitrary integer $N$, one can construct many integers $N=N'+K$,
such that $N'=n_1n_2$, where $n_1$ and $n_2$ are integers. If there exists the
probability vector with $N$ components, a new probability vector with $N'$
components can be constructed, and the $K$ components of the constructed
vector can be assumed as zero components. Then the numbers $1,2,\ldots,N'$ can
be mapped onto pairs of integers
$(1,1),(1,2),\ldots,(1,n_2),(2,1),(2,2),\ldots,(2,n_2),\ldots,(n_1,1),(n_1,2),\ldots,(n_1,n_2)$.
This means that the probability vector constructed is mapped onto a matrix
with matrix elements analogous to the joint probability distribution of two
random variables. In view of the known subadditivity condition for this joint
probability distribution, one has the entropic inequality, which can be
expressed in terms of components of the initial probability $N$-vector. We
used such a procedure to obtain the both classical and quantum strong
subadditivity conditions.

The physical interpretation of the obtained strong subadditivity condition
needs an extra clarification. There is a possibility to connect the new
entropic inequalities with such state characteristics as purity or such
parameters as Tr$\,\hat\rho^n$, as well as with correlations between different
groups of measurable quantities. The entropic interpretation can be given to
the correlations between groups of the tomographic-probability values.

The tomographic distributions and their relations to different
quasidistributions obtained in \cite{JPA-Rui} can be used to derive entropic
inequalities associated with analytic signals.

New relations for $q$-entropies obtained for multipartite systems in
\cite{MVJRLR} and associated with entropic inequalities discussed in \cite{7}
can be also considered for systems without subsystems, in view of the approach
developed. We apply this procedure to find new equalities and inequalities for
probability distributions and density matrices of quantum states in a future
publication.

\section*{Appendix}
We present the entropic inequalities -- the subadditivity conditions for the
spin tomographic probability distributions $w(m,\vec n)$ for one qudit with
spin $j=2$ and $j=3$ as follows:

$j=2$,  $~m=-2,-1,0,1,2$, and $~\vec
n=(sin\theta\cos\varphi,sin\theta\sin\varphi,cos\theta)$,
\begin{eqnarray*}
\fl -\big[w(-2,\vec n)+w(-1,\vec n)+w(0,\vec n)\big]\ln\big[w(-2,\vec
n)+w(-1,\vec n)+w(0,\vec n)\big]\nonumber\\
\fl-\big[w(2,\vec n)+w(2,\vec n)\big]\ln\big[w(1,\vec n)+w(2,\vec
n)\big]-\big[w(-2,\vec n)+w(1,\vec
n)\big]\ln\big[w(-2,\vec n)+w(1,\vec n)\big]\nonumber\\
\fl -\big[w(0,\vec n)+w(2,\vec n)\big]\ln\big(w(0,\vec n)+w(2,\vec n)\big]
\nonumber\\
\fl \geq -\big[w(-2,\vec n)\ln w(-2,\vec n)+w(0,\vec n)\ln w(0,\vec
n)+w(1,\vec n)\ln w(1,\vec n)+w(2,\vec n)\ln w(2,\vec n)\big],\nonumber\\
\end{eqnarray*}

$j=3$, $~m=-3,-2,-1,0,1,2,3$, and $~\vec
n=(sin\theta\cos\varphi,sin\theta\sin\varphi,cos\theta)$,
\begin{eqnarray*}
\fl -\big[w(-3,\vec n)\ln w(-3,\vec n)+w(-2,\vec n)\ln w(-2,\vec n) +w(-1,\vec
n)\ln w(-1,\vec n)\nonumber\\
\fl  +w(1,\vec n)\ln w(1,\vec n)+w(2,\vec n)\ln w(2,\vec
n)\big]\nonumber\\
\fl -\big[w(-1,\vec n)+w(0,\vec n)+w(3,\vec n)\big]\ln\big[w(-1,\vec
n)+w(0,\vec n)+w(3,\vec
n)\big]\nonumber\\
\fl -\big[w(-3,\vec n)+w(-2,\vec n)+w(1,\vec n)+w(2,\vec
n)\big]\ln\big[w(-3,\vec n)+w(-2,\vec n)+w(1,\vec n)+w(2,\vec n)\big]
\nonumber\\
\fl \leq -\big[w(-3,\vec n)+w(-2,\vec n)\big]\ln\big[w(-3,\vec n)+w(-2,\vec
n)\big]\nonumber\\
\fl -\big[w(-1,\vec n)+w(0,\vec n)\big]\ln\big[w(-1,\vec n)+w(0,\vec n)\big]
-\big[w(1,\vec
n)+w(2,\vec n)\big]\ln\big[w(1,\vec n)+w(2,\vec n)\big]\nonumber\\
\fl -\big[w(-3,\vec n)+w(1,\vec n)\big]\ln\big[w(-3,\vec n)+w(1,\vec
n)\big]\nonumber\\
\fl -\big[w(-2,\vec n)+w(2,\vec n)\big]\ln\big[w(-2,\vec n)+w(2,\vec
n)\big]\nonumber\\
\fl -\big[w(-1,\vec n)+w(3,\vec n)\big]\ln\big[w(-1,\vec n)+w(3,\vec n)\big].
\nonumber
\end{eqnarray*}
One can easily obtain the other inequalities by arbitrary permutations of the
spin projections, i.e., the set of all $m$ can be replaced by arbitrary
permutations of the values of spin projections.

Also these inequalities can be checked experimentally.

\section*{Acknowledgements}
This study was initiated by the memory of our discussions with
Prof.~Rui~Vilela~Mendes and Prof.~Mary~Beth~Ruskai during the Madeira Math
Encounters XXVI (October 3--11, 2003) Quantum Information, Control and
Computing. We thank Rui~Vilela~Mendes for reading this manuscript before
publication and giving useful comments. This work was supported by the Russian
Foundation for Basic Research under Project No.~11-02-00456\_a. We are
grateful to the Organizers of the XX Central European Workshop on Quantum
Optics (Stockholm, Sweden, June 16--20, 2013) and especially to Prof. Gunnar
Bjork for kind hospitality.


\section*{References}

\begin{thebibliography}{99}

\bibitem{Shannon}
 Shannon C E 1948 {\it Bell Syst. Tech. J.} {\bf 27} 379

\bibitem{Bell}
Bell J 1964 {\it Physics} {\bf 1} 195

\bibitem{HornClauser}
Clauser J F, Horne M A, Shimony A, Holt R 1969 {\it Phys. Rev. Lett.} {\bf 23}
880

\bibitem{Mendes}
Man'ko M A, Man'ko V I, Vilela Mendes R 2006
{\it J. Russ. Laser Res.} {\bf 27} 507                   

\bibitem{NuovoCim}
Man'ko M A, Man'ko V I, Marmo G, Simoni A, Ventriglia F 2013 "Introduction to
tomography, classical and quantum," {\it Nuovo Cimento, Colloquia and
Communications in Physics, Mathematical Structures in Quantum Systems and
Applications} {\bf 36} Ser.~3,  p.~163

\bibitem{VovaJETP}
Man'ko O V, Chernega V N 2013 {\it JETP Lett.} {\bf 97} 557

\bibitem{LiebRuskai}
Lieb E H, Ruskai M B 1973 {\it J. Math. Phys.} {\bf 14} 1938

\bibitem{Ruskai}
Ruskai M B 2004 arXiv: quant-ph/0404126 v4

\bibitem{Lieb}
Carlen E A, Lieb E H 2008
{\it Lett. Math. Phys.} {\bf 83} 107   

\bibitem{6}
Kim I H 2011 arXiv:1210.5190

\bibitem{4}
Ohya M, Petz D 2004 {\it Quantum Entropy and Its Use} Springer/Heidelberg, 2nd
edition

\bibitem{5}
Ruskai M B 2007 Rep. Math. Phys. {\bf 60} 1 quant-ph/0604206

\bibitem{8}
Frank L R, Lieb E H 2012 arXiv:1204.0825v1 quant-ph

\bibitem{RitaFP}
 Man'ko M A, Man'ko V I 2011 {\it Found.~Phys.} {\bf 41} 330

\bibitem{VovJRLR2013}
Chernega V N, Man'ko O V 2013 {\it J. Russ. Laser Res.} {\bf 34} 383

\bibitem{portJRLR}
Chernega V N, Man'ko V I 2007 {\it J. Russ. Laser Res.} {\bf 28} 103  

\bibitem{LupoJPA}
Lupo C, Man'ko V I, Marmo G 2007 {\it J. Phys. A: Math. Theor.} {\bf 40} 13091

\bibitem{DodPLA}
Dodonov V V, Man'ko V I 1997 {\it Phys. Lett. A} {\bf 229} 335

\bibitem{OlgaJETP}
Man'ko V I, Man'ko O V 1997 {\it J. Exp. Theor. Phys.} {\bf 85} 430

\bibitem{MVJRLR}
Man'ko M A, Man'ko V I 2013 {\it J. Russ. Laser Res.} {\bf 34} 203  

\bibitem{JPA-Rui}
Man'ko M A, Man'ko V I, Vilela Mendes R 2001 {\it J.~Phys. A: Math. Gen.} {\bf
34} 8321

\bibitem{7}
Rastegin A E 2012 
arXiv:1210.6742 quant-ph







\end{thebibliography}
\end{document}